\documentclass[aps,prd,preprintnumbers,nofootinbib,superscriptaddress,floatfix,tightenlines,twocolumn,compress,10pt]{revtex4-2}
\usepackage[utf8]{inputenc}
\usepackage[english]{babel}
\usepackage{amsmath,amssymb,amsthm,amsfonts}
\usepackage{array}
\usepackage{bm}
\usepackage{multirow}
\usepackage{graphicx}
\usepackage[dvipsnames]{xcolor}
\usepackage[hypertexnames=false,colorlinks=true,citecolor=blue,urlcolor=blue,linkcolor=black]{hyperref}

\begin{document}
\preprint{INHA-NTG-09/2026}
\title{Isospin and flavor SU(3) symmetry breaking in doubly heavy 
  baryons} 
\author{Ghil-Seok Yang}
\email{ghsyang@hoseo.edu}
\affiliation{Department of Physics, Hoseo University, Asan 31499,
  Republic of Korea} 

\author{June-Young Kim}
\email{jun-young.kim@inha.ac.kr}
\affiliation{Department of Physics, Inha University, Incheon 22212,
  Republic of Korea} 
\affiliation{Institute of Quantum Science, Inha University, Incheon
  22212, Republic of Korea} 

\author{Hyun-Chul Kim}
\email{hchkim@inha.ac.kr}
\affiliation{Department of Physics, Inha University, Incheon 22212,
  Republic of Korea} 
\affiliation{Institute of Quantum Science, Inha University, Incheon
  22212, Republic of Korea}
\affiliation{School of Physics, Korea Institute for Advanced Study 
  (KIAS), Seoul 02455, Republic of Korea}

\begin{abstract}
  The recent observations of the $\Xi_{cc}^{+}$ and the
$\Omega_{cc}^{+}$ by the LHCb Collaboration complete the
ground-state triplet of doubly charmed baryons, so that its
internal mass pattern can be used to test symmetry breaking in the
light-quark sector. We study the doubly heavy baryons within the
pion mean-field approach, regarding them as bound states of the
$N_c-2$ valence light quarks that generate the pion mean field in
the presence of a compact heavy diquark. All dynamical parameters
for flavor SU(3) and isospin symmetry breaking were determined in 
the light and singly heavy baryon sectors, and the hyperfine
coupling is fixed by the singly heavy spectrum. With the heavy-quark
masses taken from the PDG values, the measured $\Xi_{cc}^{++}$ mass is 
used only to determine the collective rotational energy of the light
subsystem, so that the mass splittings within the triplet are
predictions. We obtain $M_{\Xi_{cc}^{+}}=(3620.17\pm13.08)$ MeV and 
$M_{\Omega_{cc}^{+}}=(3746.62\pm13.22)$ MeV. The predicted isospin
splitting, $M_{\Xi_{cc}^{++}}-M_{\Xi_{cc}^{+}}=(1.43\pm0.38)$ MeV,
is consistent with the LHCb observation of the $\Xi_{cc}^{+}$. We
also present the masses of the spin-3/2 doubly charmed states and
the doubly bottom spectrum.
\end{abstract}

\maketitle
%
% \tableofcontents
%%%%%%%%%%%%%%%%%%%
\section{Introduction}
\label{sec:1}
%%%%%%%%%%%%%%%%%%% 
Half a century after their existence was first anticipated in the
quark model~\cite{DeRujula:1975qlm,Gaillard:1974hs}, the
ground-state doubly charmed baryons have all been observed. The
$\Xi_{cc}^{++}$ was established by the LHCb Collaboration in 2017 in
the $\Lambda_c^+K^-\pi^+\pi^+$ final state~\cite{LHCb:2017xicc}, and
its mass was subsequently measured with high
precision~\cite{LHCb:2019epo}. Earlier this year, its isospin partner
$\Xi_{cc}^{+}$ was observed in the $\Lambda_c^+K^-\pi^+$ final state
with the LHCb Run~3 detector~\cite{LHCb:2026Xiccplus}. Most recently,
the observation of the $\Omega_{cc}^{+}$ in the $\Omega_c^0\pi^+$
mass spectrum was reported at the Beauty 2026 conference, with a
preliminary mass of
$(3725.9\pm1.2)~\mathrm{MeV}$~\cite{LHCb:2026Omegacc}.  
The complete spin-$1/2$ triplet now makes it possible to test
different symmetry-breaking mechanisms separately.
The new observations also settle a long-standing experimental
controversy.

The $\Xi_{cc}^{+}$ was first reported by the SELEX
Collaboration at $(3518.9\pm0.9)~\mathrm{MeV}$~\cite{SELEX:2002wqn,
  SELEX:2004lln}, but no state at that mass was seen in subsequent
searches by BABAR~\cite{BaBar:2006bab}, Belle~\cite{Belle:2006edu,
  Belle:2013htj}, and LHCb~\cite{LHCb:2013hvt, LHCb:2019gqy,
  LHCb:2021eaf}. Both the SELEX candidate and the $\Xi_{cc}^{++}$ decay
weakly and would thus belong to the same ground-state multiplet, so
that their mass difference of about $100~\mathrm{MeV}$ would have to
be an isospin splitting, whereas isospin splittings of heavy baryons
are expected to be at most a few MeV~\cite{Karliner:2019lau}. The
$\Xi_{cc}^{+}$ mass measured by LHCb is consistent with the expected
isospin splitting. An earlier LHCb search for the $\Omega_{cc}^{+}$
likewise found no significant signal~\cite{LHCb:2021Omegacc}, and only
the recent observation completes the triplet. 

Theoretical studies of doubly heavy baryons began long before their
observation. Early work based on heavy-quark symmetry related
their excitation spectra to those of heavy mesons through the
correspondence between a heavy diquark and a heavy
antiquark~\cite{Savage:1990di}. Their masses were computed in various
approaches such as nonrelativistic and relativistic quark
models~\cite{Ebert:2002ig,Roberts:2007ni,Karliner:2014gca}, QCD sum
rules together with potential models~\cite{Kiselev:2001fw}, and
lattice QCD~\cite{Namekawa:2013vu, Alexandrou:2014sha, Brown:2014ena,
  Mathur:2018rwu}, and the isospin splittings of heavy baryons,
including the doubly charmed states, were investigated in
Refs.~\cite{Hwang:1986ee,Hwang:2008dj,Brodsky:2011zs,
  Karliner:2017gzy, Karliner:2019lau}. The new observations have
triggered renewed theoretical interest in the lifetimes of doubly
heavy baryons~\cite{Cheng:2026dhb,Dulibic:2026dhb} and in their
possible excited states~\cite{Wang:2026dcb}.

In Ref.~\cite{Yang:2016qdz}, a singly heavy baryon was described as
a bound state of the $N_c-1$ light valence quarks that produce the
pion mean field, with the heavy quark treated as a static color
source. This description is rooted in Witten's large-$N_c$ picture, in
which a light baryon is viewed as a bound state of $N_c$ valence
quarks in a mean field~\cite{Witten:1979kh,Witten:1983}. The
presence of the $N_c$ valence quarks polarizes the Dirac sea, and
the vacuum polarization in turn influences the valence quarks. This
self-consistent process yields the pion mean
field~\cite{Diakonov:1987ty, Christov:1995vm, Diakonov:1997sj}. With
the pion mean field 
generated by the $N_c-1$ valence quarks, this approach reproduced the
masses of the lowest-lying singly heavy baryons and was subsequently
applied to various properties of these baryons such as the mass
spectra with flavor SU(3) and isospin symmetry
breaking~\cite{Kim:2018xlc, Kim:2019orc, Yang:2020klp}, the magnetic 
moments~\cite{Yang:2018uoj}, the electromagnetic form
factors~\cite{Kim:2018nqf,Kim:2019khk}, the magnetic transitions and
radiative decays~\cite{Yang:2019dzb,Kim:2021emtrans}, and the medium 
modification of the masses~\cite{Ghim:2022medium}.

Replacing one more light quark with a heavy quark leads to a doubly
heavy baryon. It contains a single light quark and thus belongs to
the flavor triplet, while the two heavy quarks form a compact
color-antitriplet diquark in the limit of $m_Q\to\infty$, which
acts as a static color source~\cite{Falk:1993xicc}. In the present
work, we extend the approach of Ref.~\cite{Yang:2016qdz} to doubly
heavy baryons, regarding them as bound states of the $N_c-2$ light
valence quarks that generate the pion mean field in the presence of
the compact heavy diquark. Although the $N_c-2$ valence quarks
polarize the vacuum less strongly than the $N_c$ and $N_c-1$ valence
quarks, the polarization is still strong enough to produce the pion
mean field. 

We emphasize that the present analysis does not require solving the
pion mean field self-consistently for the $N_c-2$ valence quarks.
Following the model-independent approach~\cite{Adkins:1984tr,
Yang:2010fm}, we retain only the collective structure dictated by
the hedgehog symmetry and the embedding into flavor SU(3), and all
dynamical coefficients are fixed by the experimental data. As will
be shown in the present work, this weaker mean field describes the
mass spectrum of the doubly heavy baryons consistently. The same
framework also reproduces their decay widths well, which will be
presented in a separate publication.

The three ground states share the same compact $cc$ core, so that
their mass differences are governed primarily by the light degrees
of freedom. The $\Xi_{cc}^{++}(ccu)$--$\Xi_{cc}^{+}(ccd)$ splitting
probes isospin breaking, which arises from the $u$--$d$ quark-mass
difference and electromagnetic effects. The
$\Xi_{cc}^{+}(ccd)$--$\Omega_{cc}^{+}(ccs)$ splitting instead
replaces $d$ by $s$ while keeping both the heavy core and the total
electric charge unchanged, and thus provides a direct measure of
the flavor SU(3) breaking effects on the light subsystem. To describe
these mass differences, we employ the 
collective Hamiltonian for isospin and flavor SU(3) symmetry
breaking developed for the SU(3) light
baryons~\cite{Yang:2010fm,Yang:2010id} and already applied to the
singly heavy baryons~\cite{Yang:2016qdz,Yang:2020klp}. Since all
dynamical parameters were determined in these sectors, the
present extension to the $N_c-2$ case introduces no additional free
parameters. Even the strength of the hyperfine interaction that
separates the spin-1/2 and spin-3/2 doubly heavy baryons is fixed
by the singly heavy spectrum. The central question of the present
work is whether this description, with no parameters adjusted, can
account for both observed splittings simultaneously.

The present paper is organized as follows. In Sec.~\ref{sec:2}, we
formulate the pion mean-field description of the light subsystem in
the presence of a compact heavy diquark and derive the
contributions of the collective rotation, the hyperfine
interaction, and the flavor SU(3) and isospin symmetry breakings to
the masses of the doubly heavy baryons. In Sec.~\ref{sec:3}, we first
explain how the collective rotational energy of the light subsystem is
determined by the measured $\Xi_{cc}^{++}$ mass, and then examine the
effects of isospin and flavor SU(3) breaking in comparison with the
observed doubly charmed spectrum. We also present the masses of the
spin-3/2 doubly charmed states and the doubly bottom spectrum. The
final section is devoted to the summary and outlook.
In Appendix A, the mass spectra of the singly heavy baryons are
reanalyzed, from which the hyperfine coupling is determined. The
agreement obtained in the singly heavy
sector~\cite{Yang:2016qdz, Yang:2020klp} is confirmed with the
updated experimental data, which places the present extension on a
firm footing.

%%%%%%%%%%%%%%%%%%%%%%%%%%%%%%%%%%%%%%%%%%%%%%%%%%
\section{Pion mean fields for doubly heavy baryons}
 \label{sec:2}
%%%%%%%%%%%%%%%%%%%%%%%%%%%%%%%%%%%%%%%%%%%%%%%%%%
In the limit of $m_Q\to\infty$, the compact heavy diquark acts as a
static color source and is thus decoupled from the light subsystem
described by the pion mean field. The heavy diquark affects the
light subsystem only through the hyperfine interaction of order
$1/m_Q$ and the Coulomb interaction, which will be introduced
below. We express the effective Hamiltonian as
\begin{align}
H
=
H_{QQ}^{(0)}
+H_{\mathrm{rot}}
+H_{\mathrm{hf}}
+H_{\mathrm{sb}}^{\mathrm{flavor}}
+H_{\mathrm{iso}}^{(Q)},
\label{eq:master-H}
\end{align}
where $H_{QQ}^{(0)}$ represents the common mass contribution of the
compact heavy-diquark core for a fixed heavy flavor $Q$,
$H_{\mathrm{rot}}$ describes the collective rotation of the light
subsystem bound by the pion mean field, $H_{\mathrm{hf}}$ couples
the heavy-diquark spin to the spin of the light subsystem,
$H_{\mathrm{sb}}^{\mathrm{flavor}}$ generates flavor SU(3) breaking,
and $H_{\mathrm{iso}}^{(Q)}$ contains all isospin-breaking
effects~\cite{Yang:2020klp}. Taking the expectation value of
Eq.~\eqref{eq:master-H} in a baryon state $B_{QQ,J}$ belonging to
the SU(3) representation $\mathcal R$, we obtain the mass formula
\begin{align}
M_{B_{QQ,J}}
=
2m_Q
+E_{\mathcal R}^{\mathrm{rot}}
+\Delta_{\mathrm{hf}}^{(Q,J)}
+\delta_{\bm 3}^{Y}\langle Y\rangle_B
+\Delta_{B}^{\mathrm{iso},Q}.
\label{eq:master-mass}
\end{align}
Here $\langle H_{QQ}^{(0)}\rangle=2m_Q$ is common to a fixed heavy
flavor, while the remaining four quantities are defined explicitly
in the following subsections. The rotational energy
$E_{\mathcal R}^{\mathrm{rot}}$ arises from the collective
quantization of the pion mean field, and the hyperfine shift
$\Delta_{\mathrm{hf}}^{(Q,J)}$ from the spin coupling between the
heavy diquark and the light subsystem, which yields the total spin
$J$ of the baryon $B_{QQ,J}$. The term
$\delta_{\bm 3}^{Y}\langle Y\rangle_B$ arises from explicit
flavor SU(3) breaking, and $\Delta_{B}^{\mathrm{iso},Q}$ from the
complete isospin-breaking Hamiltonian. Throughout this section,
$|B\rangle$ denotes a physical doubly heavy baryon state, with the
heavy flavor, total spin, SU(3) representation, and other quantum
numbers understood unless they need to be displayed explicitly.

%%%%%%%%%%%%%%%%%%%%%%%%%%%%%%
\subsection{Collective Hamiltonian}
\label{sec:2a}
%%%%%%%%%%%%%%%%%%%%%%%%%%%%%%
\begin{figure*}[t]
    \centering
    \includegraphics[width=\textwidth]{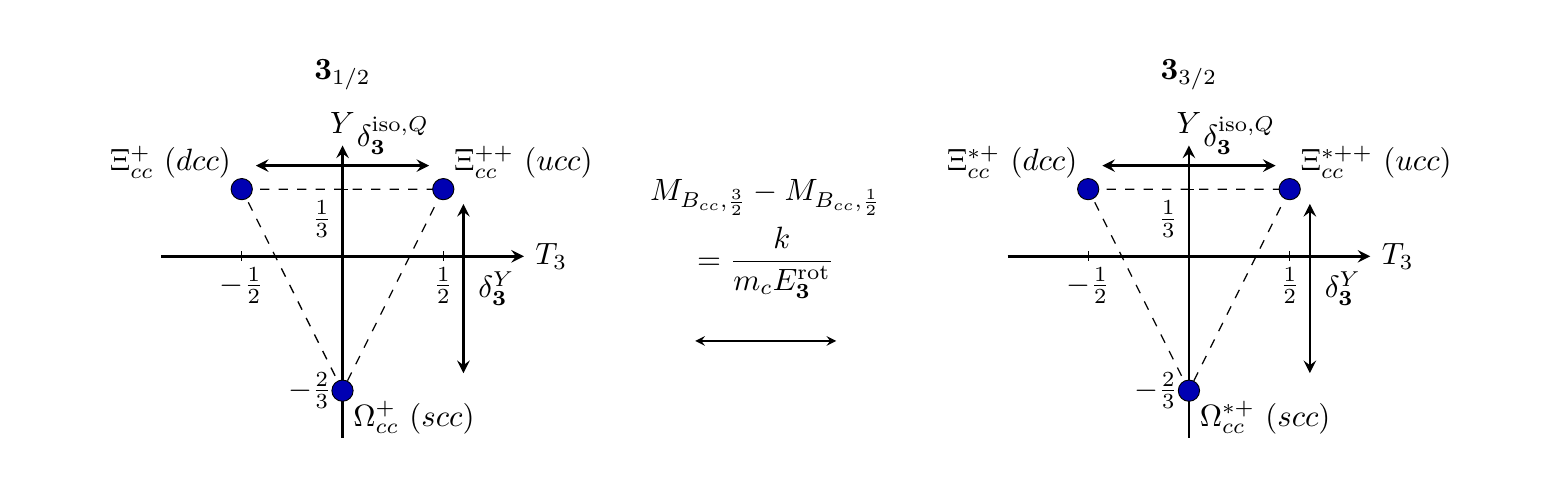}
    \caption{Weight diagrams for the lowest-lying doubly charmed
  baryons in the SU(3) triplet ($\bm{3}$) representation with spin
  1/2 (left) and 3/2 (right). The arrows indicate the hyperfine,
  flavor SU(3), and isospin splittings.}
    \label{fig:WD}
\end{figure*}
The pion mean field satisfies the hedgehog
symmetry~\cite{Skyrme:1961vq} and is embedded into flavor
SU(3)~\cite{Witten:1983}. Since the SU(3) zero-mode quantization in
the pion mean-field approach was described in detail in previous
works~\cite{Christov:1995vm,Diakonov:1997sj,Kim:2018xlc}, we start
directly from the collective Hamiltonian:
\begin{align}
H_{\mathrm{rot}}
\equiv
H_{(m,n)}^{\mathrm{rot}}
=
M_{\mathrm{cl}}
+\frac{1}{2I_1}\sum_{i=1}^{3}\hat J_i^2
+\frac{1}{2I_2}\sum_{a=4}^{7}\hat J_a^2,
\end{align}
where $M_{\mathrm{cl}}$ is the classical mass of the pion mean-field
configuration, $I_{1,2}$ are the moments of inertia, and $\hat J_a$
are the right SU(3) generators, with $\hat J_{1,2,3}$ identified with
the spin operators of the light subsystem. The collective baryon wave
functions are given by the SU(3) Wigner $D$ functions, from which the
rotational eigenvalue is obtained as
\begin{align}
E_{\mathcal R}^{\mathrm{rot}}
&=
M_{\mathrm{cl}}
+\frac{J_{L}(J_L+1)}{2I_1} \nonumber \\
&+\frac{C_2(m,n)-J_L(J_L+1)-\frac{3}{4}Y^{\prime\,2}}{2I_2},
\end{align}
where $C_2(m,n)$ is the quadratic Casimir of the SU(3) representation
$\mathcal{R}=(m,n)$, and $J_L$ denotes the spin of the
light subsystem.

The constraint on $Y'$ selects the tower of allowed rotational
excitations. For the light baryons, the $N_c$ valence quarks fix
$Y'=N_c/3$, which selects the octet and the decuplet as the lowest
allowed representations, in exact correspondence with the quark
model. For singly heavy baryons, the $N_c-1$ valence light quarks
modify the constraint to $Y'=(N_c-1)/3$, and the lowest allowed
representations become the antitriplet and the
sextet~\cite{Yang:2016qdz}. In the doubly heavy sector, the $N_c-2$
valence light quarks correspondingly yield
\begin{equation}
Y'=\frac{N_c-2}{3}.
\end{equation}
Since the states with hypercharge $Y'$ must have isospin equal to
the spin $J_L$ of the light subsystem, the lowest allowed
representation is the triplet,
\begin{equation}
\mathcal R=\bm{3}=(1,0),
\end{equation}
with $J_L=1/2$. This again agrees with the quark model, in which
the single light quark in a doubly heavy baryon belongs to the
flavor triplet. For this representation, the rotational energy of
the light subsystem is given by
\begin{align}
E_{\mathcal R}^{\mathrm{rot}} &=  M_{\mathrm{cl}} + \frac{3}{8I_1} +
    \frac{1}{4I_2},
  &&\left(\mathcal{R}=\bm{3}, J_L=1/2\right). 
\label{eq:rot}
\end{align}
%%%%%%%%%%%%%%%%%%%%%%%%%%%%%%%%%%%%%%%%%%
\subsection{Spin structure and hyperfine splitting}
\label{sec:2b}
%%%%%%%%%%%%%%%%%%%%%%%%%%%%%%%%%%%%%%%%%%
Equation~\eqref{eq:rot} determines the rotational energy of the
light subsystem. A doubly heavy baryon is obtained by coupling this
subsystem to the heavy diquark. In the infinitely heavy-quark mass
limit, this heavy diquark acts as a static color source. To form an
overall color singlet together with the light quark, the heavy
quark pair must be in the color-antisymmetric configuration. For
the ground state, the spatial wave function of the heavy quark pair
is symmetric, and the two heavy quarks carry the same flavor. The
spin wave function of the pair is therefore symmetric, which fixes
the heavy-diquark spin to be
\begin{equation}
\bm J_H=\bm J_{h_1}+\bm J_{h_2},
\qquad J_H=1.
\end{equation}
Coupling the heavy-diquark spin to the spin of the light subsystem
with $J_L=1/2$ yields the two lowest-lying states
\begin{equation}
\bm J=\bm J_H+\bm J_L,
\qquad J=\frac12,\,\frac32,
\end{equation}
which are degenerate in the limit of $m_Q\to\infty$. The degeneracy
is lifted by the hyperfine interaction between the heavy diquark
and the light subsystem,
\begin{align}
H_{\mathrm{hf}}
&=
\frac{2}{3}\frac{k}{m_Q E_{\mathcal R}^{\mathrm{rot}}}
\bm J_H\cdot\bm J_L,
\label{eq:HhfDHB}
\end{align}
where the hyperfine coupling constant $k$ is fixed in the singly
heavy sector, and $E_{\mathcal R}^{\mathrm{rot}}$ denotes the
rotational energy of the light subsystem in the SU(3)
representation $\mathcal R$. For the lowest-lying doubly heavy
states considered here, we have $\mathcal R=\bm 3$. Thus, the same
light-subsystem energy that appears in the spin-independent mass
formula also sets the mass scale of the hyperfine interaction. The
spin-spin interaction between the two heavy quarks generates a
common shift for the fixed ground-state diquark with $J_H=1$, so
that it does not affect the splitting between the $J=1/2$ and
$J=3/2$ states. The hyperfine contribution in
Eq.~\eqref{eq:master-mass} is then defined by
\begin{align}
\Delta_{\mathrm{hf}}^{(Q,J)}
&\equiv
\langle B|H_{\mathrm{hf}}|B\rangle \nonumber \\
&=
\frac{1}{3}\frac{k}{m_Q E_{\mathcal R}^{\mathrm{rot}}} \nonumber \\
&\times \left[
J(J+1)-J_H(J_H+1)-J_L(J_L+1)
\right].
\label{eq:Dhf-general}
\end{align}
For $J_H=1$ and $J_L=1/2$, we obtain
\begin{subequations}
\begin{align}
\Delta_{\mathrm{hf}}^{(Q,1/2)}
&=-\frac{2}{3}\frac{k}{m_Q E_{\mathcal R}^{\mathrm{rot}}},
\\
\Delta_{\mathrm{hf}}^{(Q,3/2)}
&=+\frac{1}{3}\frac{k}{m_Q E_{\mathcal R}^{\mathrm{rot}}}.
\end{align}
\end{subequations}
The splitting between different spin states is then given by
\begin{align}
M_{B_{QQ,3/2}}-M_{B_{QQ,1/2}}
&=
\Delta_{\mathrm{hf}}^{(Q,3/2)}
-\Delta_{\mathrm{hf}}^{(Q,1/2)} \nonumber \\
&=
\frac{k}{m_Q E_{\mathcal R}^{\mathrm{rot}}}.
\label{eq:hfsplitDHB}
\end{align}
The numerical value of $k$ was already fixed by the hyperfine
splittings of the charmed singly heavy baryons, as described in
Appendix~\ref{app:a}. The $1/m_Q$ dependence in
Eq.~\eqref{eq:hfsplitDHB} then naturally suppresses the hyperfine
splitting in the bottom sector.

%%%%%%%%%%%%%%%%%%%%%%%%%%%%%%%%%%%%%%
\subsection{Flavor SU(3) symmetry breaking}
\label{sec:2c}
%%%%%%%%%%%%%%%%%%%%%%%%%%%%%%%%%%%%%%
The rotational and hyperfine terms determine the spectrum in the
flavor SU(3) symmetric limit. Explicit flavor SU(3) symmetry
breaking by the strange current-quark mass then separates the
$\Xi_{QQ}$ and $\Omega_{QQ}$ states along the hypercharge axis $Y$,
as illustrated in Fig.~\ref{fig:WD}. Treating the difference
between the strange-quark mass $m_s$ and the average light-quark
mass $\bar m=(m_u+m_d)/2$ perturbatively, we write the Hamiltonian
for flavor SU(3) symmetry breaking as
\begin{align}
H_{\mathrm{sb}}^{\mathrm{flavor}}
&=
\left(m_{\mathrm s}-\bar m\right) \nonumber \\
&\times\left(
\bar{\alpha}\, D_{88}^{(8)}
+\beta \hat{Y}
+\frac{\gamma}{\sqrt{3}}\sum_{i=1}^{3} D_{8i}^{(8)} \hat{J}_i
\right).
\label{eq:sb_H}
\end{align}
Here, $D_{ab}^{(8)}(A)$ denotes an SU(3) Wigner $D$ function of the
collective rotation $A$, and the argument $A$ will be suppressed
below. The dynamical parameters $\alpha$, $\beta$, and $\gamma$ are
fixed by the light-baryon spectrum. Their application to doubly
heavy baryons requires one additional assumption that the pion mean
field is now generated by the $N_c-2$ valence light quarks rather
than by the $N_c$ ones. Following the treatment of singly heavy
baryons in Ref.~\cite{Yang:2016qdz}, we account for this change by
rescaling the coefficient $\alpha$ as
\begin{align}
\alpha \to \bar{\alpha} = \frac{N_c-2}{N_c} \alpha,
\end{align}
while keeping $\beta$ and $\gamma$ unchanged. The bar distinguishes
the modified coefficient from its counterpart in the light-baryon
sector. Using the light-baryon parameters of
Ref.~\cite{Yang:2010fm}, this prescription yields
\begin{subequations}
\label{eq:ABR}
\begin{align}
\left(m_{\mathrm{s}}-\bar{m}\right)\bar{\alpha}
& =(-85.01\pm1.94)\;\mathrm{MeV}\,, \\[0.5ex]
\left(m_{\mathrm{s}}-\bar{m}\right)\beta
& =(-140.04\pm3.20)\;\mathrm{MeV}\,, \\[0.5ex]
\left(m_{\mathrm{s}}-\bar{m}\right)\gamma
& =(-101.08\pm2.33)\;\mathrm{MeV}\,.
\end{align}
\end{subequations}
Evaluating the matrix elements of Eq.~\eqref{eq:sb_H} in the triplet
states, we obtain the fourth term of Eq.~\eqref{eq:master-mass},
\begin{align}
\langle B|H_{\mathrm{sb}}^{\mathrm{flavor}}|B\rangle
=
\delta_{\bm{3}}^{Y}\langle Y\rangle_B,
\label{eq:flavor-matrix}
\end{align}
where $\langle Y\rangle_B\equiv\langle B|\hat Y|B\rangle$ is the
expectation value of the hypercharge in the baryon state $B$. The
coefficient $\delta_{\bm{3}}^{Y}$ is given by
\begin{align}
\delta_{\bm{3}}^{Y}
&=
\frac{3}{16}(m_s-\bar{m})
\left(
\bar{\alpha}
+\frac{16}{3}\beta
-\frac{3}{2}\gamma
\right) \nonumber \\[1ex]
&= (-127.55 \pm 3.28)~\mathrm{MeV}.
\end{align}
Neglecting the isospin-breaking term $\Delta_{B}^{\mathrm{iso},Q}$, 
Eq.~\eqref{eq:master-mass} leads to 
\begin{align}
M_{\Xi_{QQ,J}}
&=
2m_Q+E_{\mathcal R}^{\mathrm{rot}}
+\Delta_{\mathrm{hf}}^{(Q,J)}
+\frac{1}{3}\delta_{\bm{3}}^{Y},
\\
M_{\Omega_{QQ,J}}
&=
2m_Q+E_{\mathcal R}^{\mathrm{rot}}
+\Delta_{\mathrm{hf}}^{(Q,J)}
-\frac{2}{3}\delta_{\bm{3}}^{Y}.
\label{eq:mass-before-IB}
\end{align}
Here the isospin multiplets remain degenerate, so that the electric
charges are not displayed explicitly. The corresponding flavor
SU(3) mass splitting is then given by
\begin{align}
M_{\Xi_{QQ,J}} - M_{\Omega_{QQ,J}} = \delta_{\bm{3}}^{Y}.
\end{align}

%%%%%%%%%%%%%%%%%%%%%%%%%%%%%%
\subsection{Isospin mass splittings}
\label{sec:2d}
%%%%%%%%%%%%%%%%%%%%%%%%%%%%%%
Flavor SU(3) symmetry breaking separates states with different
strangeness but leaves the members of each isospin multiplet
degenerate. The final term in Eq.~\eqref{eq:master-mass} is
generated by the complete isospin-breaking Hamiltonian
\begin{align}
H_{\mathrm{iso}}^{(Q)}
=
H_{\mathrm{sb}}^{\mathrm{iso}}
+\mathcal O^{\mathrm{EM}}
+H_{\mathrm{Coul}}^{(Q)},
\label{eq:Hiso}
\end{align}
which contains, respectively, the effect of the $u$--$d$ quark-mass
difference, the electromagnetic self-energy, and the Coulomb
interaction between the light subsystem and the heavy quarks. In
contrast to the singly heavy baryons, where the strong hyperfine
interaction between the two light valence quarks contributes to the
isospin mass differences~\cite{Yang:2020klp}, no such contribution
arises here, since only a single light valence quark remains in a
doubly heavy baryon. We define the corresponding state-dependent
correction by
\begin{align}
\Delta_{B}^{\mathrm{iso},Q}
\equiv
\langle B|H_{\mathrm{iso}}^{(Q)}|B\rangle.
\label{eq:Diso}
\end{align}
These three contributions are evaluated below.

\medskip
\noindent\textit{Hadronic contribution.---}
The mass difference between the $u$ and $d$ quarks generates the
strong-interaction part of the isospin splitting. The corresponding
collective Hamiltonian is given by
\begin{align}
H_{\mathrm{sb}}^{\mathrm{iso}}
&=
\left(m_{\mathrm{d}}-m_{\mathrm{u}}\right) \nonumber \\
&\quad \times \left(\frac{\sqrt{3}}{2}\bar{\alpha}
D_{38}^{(8)}
+ \beta\hat{T}_{3}
+
\frac{1}{2}\gamma\sum_{i=1}^{3}D_{3i}^{(8)}\hat{J}_{i}\right)\,.
\label{eq:Hhad_iso}
\end{align}
The same dynamical parameters that govern flavor SU(3) symmetry
breaking in Eq.~\eqref{eq:sb_H} also determine this contribution.
In particular, the coefficient $\bar\alpha$ carries the same
valence-content rescaling as in Eq.~\eqref{eq:sb_H}. Multiplied by
the $u$--$d$ quark-mass difference, the parameters are given as
\begin{align}
\left(m_{\mathrm{d}}-m_{\mathrm{u}}\right)\overline{\alpha}
& =(-1.47\pm0.001)\;\mathrm{MeV}\,,\nonumber \\[0.5ex]
\left(m_{\mathrm{d}}-m_{\mathrm{u}}\right)\beta
& =(-2.41\pm0.001)\;\mathrm{MeV}\,,\nonumber \\[0.5ex]
\left(m_{\mathrm{d}}-m_{\mathrm{u}}\right)\gamma
& =(-1.74\pm0.006)\;\mathrm{MeV}\,.
\label{eq:abr}
\end{align}
Evaluating Eq.~\eqref{eq:Hhad_iso} in the triplet states, we obtain
the hadronic part of Eq.~\eqref{eq:Diso},
\begin{align}
\langle B|H_{\mathrm{sb}}^{\mathrm{iso}}|B\rangle
=
\delta^{\mathrm{iso}}_{\bm{3}}\langle T_3\rangle_B,
\label{eq:hadronic-IB}
\end{align}
where $\langle T_3\rangle_B\equiv\langle B|\hat T_3|B\rangle$ is the
expectation value of the third component of the isospin in the
baryon state $B$. The coefficient $\delta_{\bm{3}}^{\mathrm{iso}}$
is expressed as
\begin{align}
\delta_{\boldsymbol{3}}^{\mathrm{iso}}
&=\;\;\frac{3}{16}
\left(m_{d}-m_{u}\right)\left(\overline{\alpha}
+ \frac{16}{3}\beta
- \frac{3}{2}\gamma\right)\; \nonumber \\[1ex]
&=\;(-2.20\pm0.01)\,\mathrm{MeV}\,.
\label{eq:deltaiso}
\end{align}

\medskip
\noindent\textit{Electromagnetic self-energy.---}
The second contribution arises from the electromagnetic
self-energy. In the Cottingham formulation, the electromagnetic
correction to a baryon mass is written as~\cite{Cottingham:1963}
\begin{align}
M_{B}^{\mathrm{EM}}
&=\frac{1}{2}\int d^{3}x\,d^{3}y\,
\langle B|T[J_{\mu}(\bm{x})J^{\mu}(\bm{y})]|B\rangle
D_{\gamma}(\bm{x},\bm{y}) \nonumber \\
&=\langle B|\mathcal{O}^{\mathrm{EM}}|B\rangle,
\label{eq:corr}
\end{align}
where $J^{\mu}$ is the electromagnetic current defined as
\begin{align}
J^{\mu}(x)=e\,\bar{\psi}(x)\gamma_{\mu}\hat{\mathcal{Q}}\psi(x)\,.
\label{eq:EM_curr}
\end{align}
Here, $e$ denotes the elementary electric charge and
$\hat{\mathcal{Q}}$ the quark-charge operator
\begin{align}
\hat{\mathcal{Q}}
=\begin{pmatrix}
2/3 & 0 & 0\\
0 & -1/3 & 0\\
0 & 0 & -1/3
\end{pmatrix}
=\frac{1}{2}\left(\lambda_{3}+\frac{1}{\sqrt{3}}\lambda_{8}\right)\,,
\label{eq:GN}
\end{align}
where $\lambda_{3}$ and $\lambda_{8}$ are the Gell-Mann matrices of
SU(3). $D_{\gamma}$ stands for the static photon propagator, whose
effect is absorbed into phenomenological parameters. Since the
electromagnetic current transforms as a flavor octet operator, the
most general form of $\mathcal{O}^{\mathrm{EM}}$ can be written
as~\cite{Yang:2010id}
\begin{align}
\mathcal{O}^{\mathrm{EM}}
&=\alpha_{1}\sum_{i=1}^{3}D_{Qi}^{(8)}D_{Qi}^{(8)}
+\alpha_{2}\sum_{p=4}^{7}D_{Qp}^{(8)}D_{Qp}^{(8)} \nonumber \\
&\quad+\alpha_{3}D_{Q8}^{(8)}D_{Q8}^{(8)}\,,
\label{eq:emop}
\end{align}
where $D_{Qa}^{(8)}$ denotes a combination of the SU(3) Wigner $D$
functions defined by
\begin{align}
D_{Qa}^{(8)}=\frac{1}{2}\left(D_{3a}^{(8)}+\frac{1}{\sqrt{3}}D_{8a}^{(8)}\right).
\label{eq:DQa}
\end{align}
The parameters $\alpha_i$ encode the model-dependent dynamics of the
pion mean field and include the effect of the static photon
propagator. They were already fixed by the empirical electromagnetic
mass splittings of the baryon octet~\cite{Yang:2010id}.

The product of two octet operators can be decomposed into the SU(3)
irreducible representations according to
\[
\mathbf{8}\otimes\mathbf{8}
=
\mathbf{1}\oplus\mathbf{8}_{s}\oplus\mathbf{8}_{a}
\oplus\mathbf{10}\oplus\mathbf{\overline{10}}\oplus\mathbf{27}.
\]
Because of Bose symmetry, only the symmetric representations
survive, namely $\mathbf{1}$, $\mathbf{8}_{s}$, and $\mathbf{27}$,
so that $\mathcal{O}^{\mathrm{EM}}$ can be rewritten as
\begin{align}
\mathcal{O}^{\mathrm{EM}}
&=
c^{(27)}\left(
\sqrt{5}D_{\Sigma_{2}^{0}\Lambda_{27}}^{(27)}
+\sqrt{3}D_{\Sigma_{1}^{0}\Lambda_{27}}^{(27)}
+D_{\Lambda_{27}\Lambda_{27}}^{(27)}
\right) \nonumber \\
&\quad
+c^{(8)}\left(
\sqrt{3}D_{\Sigma^{0}\Lambda}^{(8)}
+D_{\Lambda\Lambda}^{(8)}
\right)
+c^{(1)}D_{\Lambda\Lambda}^{(1)}\,.
\label{eq:emop3}
\end{align}
The explicit definitions of the Wigner $D$ functions,
$D_{B_{1}B_{2}}^{(\mathcal{R})}$, can be found in
Ref.~\cite{Yang:2010fm}. The new set of parameters
$c^{(\mathcal{R})}$ can be related to the parameters $\alpha_i$,
where the superscript $(\mathcal{R})$ denotes the corresponding
irreducible representation.

The electromagnetic mass corrections to doubly heavy baryons follow
from the matrix elements of $\mathcal{O}^{\mathrm{EM}}$ between the
corresponding collective states. In principle, the coefficients
$c^{(\mathcal R)}$ may change when the mean field is modified in the
doubly heavy system. We neglect this higher-order effect here, since
the electromagnetic self-energy is already a subleading contribution
to the isospin splittings considered below.

Since the representation $\bm{3}$ does not appear in the
decomposition of $\bm{3}\otimes\bm{27}$, the $\bm{27}$-plet
contribution is absent for doubly heavy baryons in the flavor
$\bm{3}$ representation. In addition, the singlet coefficient
$c^{(1)}$ can be absorbed into the flavor-symmetric part of the mass
and therefore does not contribute to the mass splittings considered
in this work. Thus, only the octet coefficient is relevant here,
whose value was determined in Ref.~\cite{Yang:2010id} to be
\begin{align}
c^{(8)} = -0.15 \pm 0.23\,.
\end{align}

\medskip
\noindent\textit{Heavy--light Coulomb interaction.---}
The electric interaction between the light subsystem and the two
heavy quarks provides the third contribution to the isospin
splitting. The corresponding magnetic interaction is suppressed by
the heavy-quark mass~\cite{Oka:2013xxa} and will be neglected.
Assuming that the charge distribution of the light subsystem is
localized, we parametrize the Coulomb term $H_{\mathrm{Coul}}^{(Q)}$
in Eq.~\eqref{eq:Hiso} as
\begin{align}
H_{\mathrm{Coul}}^{(Q)}\equiv H_{\mathrm{sol}\mbox{-}\mathrm{h}}^{\mathrm{Coul}}
=
\alpha_{\mathrm{sol}\mbox{-}\mathrm{h}}\hat{Q}_{\mathrm{sol}}\hat{Q}_{\mathrm{h}}
+\alpha_{\mathrm{h}\mbox{-}\mathrm{h}}\hat{Q}_{\mathrm{h}}\hat{Q}_{\mathrm{h}}\,,
\label{eq:QsolQh}
\end{align}
where $\hat{Q}_{\mathrm{sol}}$ and $\hat{Q}_{\mathrm{h}}$ denote the
charge operators for the light subsystem and the heavy quarks,
respectively. The parameters $\alpha_{\mathrm{sol}\mbox{-}\mathrm{h}}$
and $\alpha_{\mathrm{h}\mbox{-}\mathrm{h}}$ contain the expectation
value of the inverse distance together with the fine-structure
constant. In practice, however, they are treated as effective
parameters and fixed by experimental data~\cite{Yang:2020klp}:
\begin{align}
\alpha_{\mathrm{sol}\mbox{-}\mathrm{h}}
=
(2.76\pm0.28)\,\mathrm{MeV}\,.
\label{eq:para}
\end{align}
As in the case of $c^{(\mathcal R)}$, possible modifications of
$\alpha_{\mathrm{sol}\mbox{-}\mathrm{h}}$ in the doubly heavy
environment are neglected. For a fixed heavy flavor, the
heavy-quark--heavy-quark term in Eq.~\eqref{eq:QsolQh} is common to
the two members of the isospin multiplet. It shifts the overall
mass but does not contribute to the isospin splitting, so that it
will be omitted.

Inserting Eqs.~\eqref{eq:hadronic-IB}, \eqref{eq:corr}, and
\eqref{eq:QsolQh} into Eq.~\eqref{eq:Diso}, we obtain the isospin
splitting of the SU(3) triplet with heavy flavor $Q$
\begin{align}
\delta_{\bm 3}^{\mathrm{iso},Q}
=
\delta_{\bm 3}^{\mathrm{iso}}
+\frac{3}{8}c^{(8)}
+2q_Q\alpha_{\mathrm{sol}\mbox{-}\mathrm{h}},
\label{eq:master-isospin}
\end{align}
where $q_Q$ denotes the electric charge of a single heavy quark in
units of $e$. For $q_c=2/3$ and $q_b=-1/3$,
Eq.~\eqref{eq:master-isospin} yields
\begin{subequations}
\begin{align}
M_{\Xi^{(*)++}_{cc}}-M_{\Xi^{(*)+}_{cc}}
&\equiv
\delta_{\bm 3}^{\mathrm{iso},c}
\nonumber \\
&=\delta_{\bm 3}^{\mathrm{iso}}
+\frac{3}{8}c^{(8)}
+\frac{4}{3}\alpha_{\mathrm{sol}\mbox{-}\mathrm{h}},
\\
M_{\Xi^{(*)0}_{bb}}-M_{\Xi^{(*)-}_{bb}}
&\equiv
\delta_{\bm 3}^{\mathrm{iso},b} \nonumber \\
&=
\delta_{\bm 3}^{\mathrm{iso}}
+\frac{3}{8}c^{(8)}
-\frac{2}{3}\alpha_{\mathrm{sol}\mbox{-}\mathrm{h}}.
\end{align}
\end{subequations}
These mass differences are generated entirely by
$\Delta_B^{\mathrm{iso},Q}$ in Eq.~\eqref{eq:master-mass}. The
common terms $2m_Q$, $E_{\mathcal R}^{\mathrm{rot}}$,
$\Delta_{\mathrm{hf}}^{(Q,J)}$, and
$\delta_{\bm 3}^{Y}\langle Y\rangle_B$ cancel within each
isodoublet.

\medskip
All the ingredients of Eq.~\eqref{eq:master-mass} are now in place.
The coefficients of flavor SU(3) and isospin symmetry breaking were
determined in the light-baryon analysis, with the valence-content
rescaling of $\bar\alpha$ described above. The heavy-quark masses
and the hyperfine coupling $k$ are fixed in the singly heavy
sector, and the electromagnetic and heavy--light Coulomb parameters
are taken from the corresponding empirical analyses. The singly
heavy spectrum obtained with these parameters is given in
Appendix~\ref{app:a}. In the doubly charmed sector, $m_c$ is not
readjusted, and the $\Xi_{cc}^{++}$ mass determines the remaining
triplet collective energy $E_{\bm 3}^{\mathrm{rot}}$. The
$\Xi_{cc}^{++}$--$\Xi_{cc}^{+}$ splitting thus tests the
isospin-breaking terms, while the
$\Xi_{cc}^{+}$--$\Omega_{cc}^{+}$ splitting tests the term for
flavor SU(3) symmetry breaking.

%%%%%%%%%%%%%%%%%%%%%%%%%%%%%%%%%%%%%%%%%%
\section{Results and discussion}
\label{sec:3}
%%%%%%%%%%%%%%%%%%%%%%%%%%%%%%%%%%%%%%%%%%
The observed doubly charmed ground states allow the framework to be
tested in a sequence that separates input, symmetry-breaking tests,
and predictions. We first specify what is fixed, then examine the
isospin and flavor SU(3) sectors in comparison with the data, and
finally turn to unobserved states.

The singly heavy calibration, including the heavy-quark masses, is
given in Appendix~\ref{app:a}. The hyperfine coupling used in the
doubly heavy sector is
\begin{align}
k=(0.10774\pm0.00071)~\mathrm{GeV}^3.
\end{align}
The remaining collective symmetry-breaking parameters were
determined in the light and singly heavy analyses described in
Sec.~\ref{sec:2}. The charm-quark mass $m_c$ is not determined
again from the doubly charmed spectrum but is taken from
Appendix~\ref{app:a}. The measured $\Xi_{cc}^{++}$ mass is the only
absolute-mass input in the doubly charmed sector and determines the
triplet collective energy $E_{\bm 3}^{\mathrm{rot}}$. With the
$J=1/2$ hyperfine shift, this condition reads
\begin{align}
&E_{\bm 3}^{\mathrm{rot}}(cc)
-\frac{2}{3}\frac{k}{m_c E_{\bm 3}^{\mathrm{rot}}(cc)}
\cr
&=M_{\Xi_{cc}^{++}}^{\mathrm{exp}}
-2m_c
-\frac{1}{3}\delta_{\bm 3}^{Y}
-\Delta_{\Xi_{cc}^{++}}^{\mathrm{iso},c}.
\label{eq:E3rot-fix}
\end{align}
Equation~\eqref{eq:E3rot-fix} determines $E_{\bm 3}^{\mathrm{rot}}$
self-consistently for the given value of $m_c$, yielding
\begin{align}
E_{\bm 3}^{\mathrm{rot}}(cc)=(1165.42\pm8.88)~\mathrm{MeV}.
\end{align}
No separate doubly charmed heavy-core mass parameter is introduced.
Once $E_{\bm 3}^{\mathrm{rot}}$ is fixed, the same value is used
for every member of the doubly charmed triplet. The
$\Xi_{cc}^{++}$--$\Xi_{cc}^{+}$ and
$\Xi_{cc}^{+}$--$\Omega_{cc}^{+}$ mass differences test,
respectively, $\Delta_{B}^{\mathrm{iso},c}$ and
$\delta_{\bm 3}^{Y}$ rather than the common collective-energy
normalization.

We first examine the isospin splitting between the $\Xi_{cc}^{++}$
and the $\Xi_{cc}^{+}$. For the spin-$1/2$ isospin partner of the
input state, we obtain
\begin{align}
M_{\Xi_{cc}^{+}}
= (3620.17\pm13.08)~\mathrm{MeV}.
\end{align}
The LHCb Run~3 observation gives
\begin{align}
M_{\Xi_{cc}^{+}}
=
\left(3619.97\pm0.83\pm0.26\,{}^{+1.90}_{-1.30}\right)
~\mathrm{MeV}
\end{align}
\cite{LHCb:2026Xiccplus}, where the first and second uncertainties
are statistical and systematic, respectively, and the third arises
from the lifetime dependence. Since the common absolute-mass
uncertainty largely cancels in the correlated mass difference, we
find
\begin{align}
M_{\Xi_{cc}^{++}}-M_{\Xi_{cc}^{+}}
=
(1.43\pm0.38)~\mathrm{MeV}.
\end{align}
Once $E_{\bm 3}^{\mathrm{rot}}$ is fixed by the $\Xi_{cc}^{++}$,
the hadronic, electromagnetic, and heavy--light Coulomb terms
reproduce the observed near-degeneracy of the isodoublet.

The strange partner tests flavor SU(3) symmetry breaking. We obtain
\begin{align}
M_{\Omega_{cc}^{+}}
= (3746.62\pm13.22)~\mathrm{MeV},
\end{align}
whereas the recent LHCb result gives
\begin{align}
M_{\Omega_{cc}^{+}}
=
(3725.9\pm1.2)~\mathrm{MeV}
\end{align}
\cite{LHCb:2026Omegacc}. The present result for $M_{\Omega_{cc}^{+}}$
is about $20.7~\mathrm{MeV}$ larger than the measured value, whereas
the lattice QCD prediction~\cite{Mathur:2018rwu} lies below the 
reported LHCb value. A similar tendency is found in the singly
heavy sector, where the same collective structure of flavor SU(3)
symmetry breaking is used: the calculated $\Omega_c$ and $\Omega_b$
masses exceed the experimental data by about $27~\mathrm{MeV}$ and
$39~\mathrm{MeV}$, respectively, as listed in
Appendix~\ref{app:a}. These deviations suggest that the
higher-order $m_{\mathrm{s}}$ corrections become relevant in the
strange sector. The second-order $m_{\mathrm{s}}$ perturbation is
known to reduce the corresponding discrepancies in the mass
splittings of the hyperons and the singly heavy
baryons~\cite{Yang:2011qe,Kim:2018xlc}. We expect that it will also
improve the description of the $\Omega_{cc}^{+}$ mass.

We now present the masses of the states that remain unobserved.
Using the hyperfine shift defined in Eqs.~\eqref{eq:Dhf-general}
and~\eqref{eq:hfsplitDHB}, we obtain the masses of the spin-$3/2$
doubly charmed states
\begin{align}
M_{\Xi_{cc}^{*++}}
&=(3694.23\pm12.51)~\mathrm{MeV},
\nonumber\\
M_{\Xi_{cc}^{*+}}
&=(3692.80\pm12.51)~\mathrm{MeV},
\nonumber\\
M_{\Omega_{cc}^{*+}}
&=(3819.25\pm12.65)~\mathrm{MeV}.
\end{align}
The corresponding $J=1/2$--$J=3/2$ hyperfine splitting is about
$(72.63\pm0.78)~\mathrm{MeV}$ and is independent of the light
flavor at the present order.

\begin{table*}[t]
\centering
\caption{Mass spectra of the lowest-lying doubly heavy baryons. The
  $\Xi_{cc}^{++}$ mass is the only absolute-mass input in the doubly
  charmed sector and determines $E_{\bm 3}^{\mathrm{rot}}$.
  Experimental values are taken from
Ref.~\cite{ParticleDataGroup:2026rpp} for $\Xi_{cc}^{++}$,
Ref.~\cite{LHCb:2026Xiccplus} for $\Xi_{cc}^{+}$, and
Ref.~\cite{LHCb:2026Omegacc} for $\Omega_{cc}^{+}$. All masses are in
MeV.} 
\label{tab:dhb-spectrum}
\renewcommand{\arraystretch}{1.25}
\begin{tabular}{c|c|c|c|c|c||c|c}
\hline
$\mathcal R_J$ & $B_{cc}$ & $Y$ & $I$ & $M_{\mathrm{th}}$ & $M_{\mathrm{exp}}$ & $B_{bb}$ & $M_{\mathrm{th}}$
\\
\hline
\multirow{3}{*}{$\bm 3_{1/2}$}
& $\Xi_{cc}^{++}$ & \multirow{2}{*}{$1/3$} & \multirow{2}{*}{$1/2$}
& $3621.60\pm13.08$ & $3621.6\pm0.4$ & $\Xi_{bb}^{0}$ &
                                                        $9899.24\pm16.11$
\\
& $\Xi_{cc}^{+}$ & &
& $3620.17\pm13.08$ & $3619.97^{+2.09}_{-1.56}$ & $\Xi_{bb}^{-}$ &
                                                        $9903.33\pm16.11$
\\
\cline{2-8}
& $\Omega_{cc}^{+}$ & $-2/3$ & $0$
& $3746.62\pm13.22$ & $3725.9\pm1.2$ & $\Omega_{bb}^{-}$ &
                                                           $10029.78\pm16.22$
\\
\hline
\multirow{3}{*}{$\bm 3_{3/2}$}
& $\Xi_{cc}^{*++}$ & \multirow{2}{*}{$1/3$} & \multirow{2}{*}{$1/2$}
& $3694.23\pm12.51$ & --- & $\Xi_{bb}^{*0}$ & $9915.50\pm16.02$
\\
& $\Xi_{cc}^{*+}$ & &
& $3692.80\pm12.51$ & --- & $\Xi_{bb}^{*-}$ & $9919.59\pm16.02$
\\
\cline{2-8}
& $\Omega_{cc}^{*+}$ & $-2/3$ & $0$
& $3819.25\pm12.65$ & --- & $\Omega_{bb}^{*-}$ & $10046.04\pm16.13$
\\
\hline
\end{tabular}
\end{table*}

In the doubly bottom sector, changing the heavy flavor from charm
to bottom suppresses $\Delta_{\mathrm{hf}}^{(Q,J)}$ through its
$1/m_Q$ dependence and changes $\Delta_{B}^{\mathrm{iso},Q}$
through the heavy-quark electric charge in
Eq.~\eqref{eq:master-isospin}. Since we do not have any experimental 
information on the mass spectrum in the doubly bottom sector, we
estimate the triplet collective energy by adding to its doubly charmed
value the average charm-to-bottom shift extracted from the antitriplet
and sextet representations of the singly heavy baryons: 
\begin{align}
E_{\bm 3}^{\mathrm{rot}}(bb)
&=E_{\bm 3}^{\mathrm{rot}}(cc)
+\frac{1}{2}\biggl[
E_{\overline{\bm 3}}^{\mathrm{rot}}(b)
-E_{\overline{\bm 3}}^{\mathrm{rot}}(c)
\nonumber\\
&\hspace{31mm}+
E_{\bm 6}^{\mathrm{rot}}(b)
-E_{\bm 6}^{\mathrm{rot}}(c)
\biggr]
\nonumber\\
&=(1582.95\pm10.60)~\mathrm{MeV}.
\end{align}
We then obtain
\begin{align}
M_{\Xi_{bb}^{0}}-M_{\Xi_{bb}^{-}}
=
(-4.09\pm0.21)~\mathrm{MeV},
\end{align}
while the hyperfine splitting is $(16.26\pm0.15)~\mathrm{MeV}$. The
resulting spectrum is summarized in Table~\ref{tab:dhb-spectrum}.

%%%%%%%%%%%%%%%%%%%%%%%%%%%%%%%%%%%%%%%%%%
\section{Summary and outlook}
\label{sec:4}
%%%%%%%%%%%%%%%%%%%%%%%%%%%%%%%%%%%%%%%%%%
In the present work, we investigated the ground-state doubly heavy
baryons within the pion mean-field approach. In the infinitely
heavy-quark mass limit, the two heavy quarks form a compact
color-antitriplet diquark that acts as a static color source, while
the dynamics of the baryon is governed by the light degrees of
freedom. Extending the description of the light and singly heavy
baryons, we regarded a doubly heavy baryon as a bound state of the
$N_c-2$ valence light quarks that generate the pion mean field in
the presence of the compact heavy diquark. The collective
quantization with the constraint $Y'=(N_c-2)/3$ selects the flavor
triplet as the lowest allowed representation, in agreement with the
quark model. All dynamical parameters for flavor SU(3) and isospin
symmetry breaking were determined in the light and singly heavy
sectors, and the hyperfine coupling was fixed by the singly heavy
spectrum. In the doubly charmed sector, only the measured
$\Xi_{cc}^{++}$ mass was used to determine the collective
rotational energy of the light subsystem, so that the mass
splittings within the triplet are predictions.

For the isospin splitting, we obtained
$M_{\Xi_{cc}^{++}}-M_{\Xi_{cc}^{+}}=(1.43\pm0.38)~\mathrm{MeV}$,
consistent with the near-degeneracy of the isodoublet observed by
LHCb. The hadronic contribution from the $u$--$d$ quark-mass
difference, the electromagnetic self-energy, and the Coulomb
interaction between the light subsystem and the heavy diquark
reproduce the data without any parameter adjusted. For the strange
partner, we obtained
$M_{\Omega_{cc}^{+}}=(3746.62\pm13.22)~\mathrm{MeV}$, which is
about $20.7~\mathrm{MeV}$ larger than the reported value, and a
similar tendency appears for the $\Omega_c$ and $\Omega_b$ in the
singly heavy sector. The second-order $m_{\mathrm{s}}$ corrections may
reduce this discrepancy. 

We also presented the masses of the spin-3/2 doubly charmed states,
with a hyperfine splitting of $72.63~\mathrm{MeV}$, and the doubly
bottom spectrum, in which the hyperfine splitting is reduced to
$16.26~\mathrm{MeV}$ by the $1/m_Q$ suppression and the isospin
splitting changes sign through the heavy-quark electric charge. We
anticipate that these predictions will soon be tested by
experiments.

\section*{Acknowledgments}
%=============================================
The present work was supported by Basic Science Research Program 
through the National Research Foundation of Korea funded by the
Ministry of Education, Science and Technology (NRF-2019R1A2C1010443 
(GSY), RS-2025-00513982 (HCK)) and by the Future-Generation
Core Researcher Joint Research Program of Inha University under the
4th Stage BK21 Graduate School Innovation Program (Project
No.~78368-8) (JYK).   
% =============================================

\appendix

%%%%%%%%%%%%%%%%%%%%%%%%%%%%%%%%%%%%%%%%%%
\section{Mass spectra of the singly heavy baryons\label{app:a}}
%%%%%%%%%%%%%%%%%%%%%%%%%%%%%%%%%%%%%%%%%%
This Appendix summarizes the parameters determined from the singly
heavy baryon sector. In the same form as the doubly heavy mass
formula in Eq.~\eqref{eq:master-mass}, the mass of a singly heavy
baryon is written as
\begin{align}
M_{B_{Q,J}}
=
m_Q
+E_{\mathcal R}^{\mathrm{rot}}
+\Delta_{\mathrm{hf}}^{(Q,J)}
+\delta_{\mathcal R}^{Y}\langle Y\rangle_B
+\Delta_{B}^{\mathrm{iso},Q},
\label{eq:SHB-master-mass}
\end{align}
where $m_Q$ is the heavy-quark mass. For a given $m_Q$,
$E_{\mathcal R}^{\mathrm{rot}}$ is determined from the singly heavy
baryon masses, while the remaining parameters for flavor SU(3) and
isospin symmetry breaking are taken from the light-baryon sector.
The heavy-quark masses are taken to be~\cite{ParticleDataGroup:2026rpp}
\begin{subequations}
\begin{align}
m_c &= (1272.9 \pm 4.5)~\mathrm{MeV},  \\[0.5ex]
m_b &= (4186.0 \pm 6.0)~\mathrm{MeV}.
\end{align}
\end{subequations}

In the charm sector, the $\Lambda_c^+$ mass determines
\begin{align}
E_{\overline{\bm 3}}^{\mathrm{rot}}(c)
&=
(1148.85\pm5.07)~\mathrm{MeV},
\label{eq:CM_ca}
\end{align}
whereas the $\Sigma_c^{++}$ and $\Sigma_c^{*++}$ masses determine
\begin{align}
E_{\bm 6}^{\mathrm{rot}}(c)
&=
(1313.54\pm5.03)~\mathrm{MeV}.
\label{eq:CM_cb}
\end{align}
The corresponding hyperfine splitting is
\begin{align}
M_{\bm 6_{3/2}}-M_{\bm 6_{1/2}}
=
\frac{k}{m_Q E_{\bm 6}^{\mathrm{rot}}},
\label{eq:hfsplitSHB}
\end{align}
which gives
\begin{align}
k
= (0.10774\pm0.00071)~\mathrm{GeV}^3.
\end{align}

The same value of $k$ is used in the bottom sector. Using the
$\Lambda_b^0$ and $\Sigma^+_b$ masses, the corresponding central
values are determined as
\begin{align}
E_{\overline{\bm 3}}^{\mathrm{rot}}(b)
&=
(1569.78\pm6.44)~\mathrm{MeV},
\\
E_{\bm 6}^{\mathrm{rot}}(b)
&=
(1727.68\pm6.39)~\mathrm{MeV}.
\label{eq:CM_b}
\end{align}
The resulting singly heavy baryon spectrum is listed in
Table~\ref{tab:shb-benchmark}.

\begin{table*}[t]
\centering
\caption{Mass spectra of the ground-state singly heavy baryons in
  comparison with the experimental
  data~\cite{ParticleDataGroup:2026rpp}. All masses and mass
  differences are in MeV. The uncertainties of the heavy-quark
  masses are included as independent inputs.}
\label{tab:shb-benchmark}
\scriptsize
\renewcommand{\arraystretch}{1.20}
\begin{tabular}{c|c|c|c|c|c|c}
\hline
$\mathcal R_J$ & $B_c$ & $Y$ & $I$ & $M_{\mathrm{th}}$ &
     $M_{\mathrm{exp}}$ & $M_{\mathrm{th}}-M_{\mathrm{exp}}$
\\
\hline
\multirow{3}{*}{$\overline{\bm 3}_{1/2}$}
& $\Lambda_c^+$ & $2/3$ & $0$ & $2286.46\pm7.18$ & $2286.46\pm0.14$ &
    input
\\
& $\Xi_c^+$ & \multirow{2}{*}{$-1/3$} & \multirow{2}{*}{$1/2$} &
       $2488.50\pm6.88$ & $2467.79\pm0.15$ & $20.71\pm6.89$
\\
& $\Xi_c^0$ & & & $2490.28\pm6.89$ & $2470.50\pm0.25$ &
                                                        $19.78\pm6.89$
\\
\hline
\multirow{6}{*}{$\bm 6_{1/2}$}
& $\Sigma_c^{++}$ & \multirow{3}{*}{$2/3$} & \multirow{3}{*}{$1$} &
  $2453.97\pm7.33$ & $2453.97\pm0.14$ & input
\\
& $\Sigma_c^+$ & & & $2453.35\pm7.32$ & $2452.65\pm0.22$ &
                                                           $0.70\pm7.33$
\\
& $\Sigma_c^0$ & & & $2454.65\pm7.32$ & $2453.75\pm0.14$ &
                                                           $0.90\pm7.32$
\\
& $\Xi_c^{\prime +}$ & \multirow{2}{*}{$-1/3$} &
    \multirow{2}{*}{$1/2$} & $2587.41\pm7.07$ & $2578.3\pm0.4$ &
                                                                   $9.11\pm7.08$
\\
& $\Xi_c^{\prime 0}$ & & & $2588.71\pm7.07$ & $2578.8\pm0.5$ &
                                                               $9.91\pm7.08$
\\
& $\Omega_c^0$ & $-4/3$ & $0$ & $2722.76\pm8.27$ & $2695.3\pm0.4$ &
                                                                    $27.46\pm8.28$
\\
\hline
\multirow{6}{*}{$\bm 6_{3/2}$}
& $\Sigma_c^{*++}$ & \multirow{3}{*}{$2/3$} & \multirow{3}{*}{$1$} &
        $2518.41\pm7.00$ & $2518.41\pm0.22$ & input
\\
& $\Sigma_c^{*+}$ & & & $2517.79\pm7.00$ & $2517.4\pm0.7$ &
                                                            $0.39\pm7.03$
\\
& $\Sigma_c^{*0}$ & & & $2519.09\pm7.00$ & $2518.48\pm0.21$ &
                                                              $0.61\pm7.00$
\\
& $\Xi_c^{*+}$ & \multirow{2}{*}{$-1/3$} & \multirow{2}{*}{$1/2$} &
    $2651.85\pm6.73$ & $2645.17\pm0.27$ & $6.68\pm6.74$
\\
& $\Xi_c^{*0}$ & & & $2653.15\pm6.73$ & $2646.24\pm0.18$ &
                                                           $6.91\pm6.73$
\\
& $\Omega_c^{*0}$ & $-4/3$ & $0$ & $2787.20\pm7.98$ & $2766.0\pm1.0$ &
                                                                       $21.20\pm8.04$
\\
\hline
\end{tabular}

\vspace{1em}

\begin{tabular}{c|c|c|c|c|c|c}
\hline
$\mathcal R_J$ & $B_b$ & $Y$ & $I$ & $M_{\mathrm{th}}$ & $M_{\mathrm{exp}}$ & $M_{\mathrm{th}}-M_{\mathrm{exp}}$
\\
\hline
\multirow{3}{*}{$\overline{\bm 3}_{1/2}$}
& $\Lambda_b^0$ & $2/3$ & $0$ & $5619.57\pm9.11$ & $5619.57\pm0.16$ &
                                                                      input
\\
& $\Xi_b^0$ & \multirow{2}{*}{$-1/3$} & \multirow{2}{*}{$1/2$} &
         $5821.61\pm8.88$ & $5791.7\pm0.4$ & $29.91\pm8.89$
\\
& $\Xi_b^-$ & & & $5826.15\pm8.88$ & $5797.0\pm0.4$ & $29.15\pm8.89$
\\
\hline
\multirow{6}{*}{$\bm 6_{1/2}$}
& $\Sigma_b^+$ & \multirow{3}{*}{$2/3$} & \multirow{3}{*}{$1$} &
    $5810.56\pm9.08$ & $5810.56\pm0.25$ & input
\\
& $\Sigma_b^0$ & & & $5812.70\pm9.08$ & --- & ---
\\
& $\Sigma_b^-$ & & & $5816.76\pm9.07$ & $5815.64\pm0.27$ &
                                                           $1.12\pm9.08$
\\
& $\Xi_b^{\prime 0}$ & \multirow{2}{*}{$-1/3$} &
   \multirow{2}{*}{$1/2$} & $5946.76\pm8.87$ & --- & ---
\\
& $\Xi_b^{\prime -}$ & & & $5950.82\pm8.87$ & $5934.9\pm0.4$ & $15.92\pm8.88$
\\
& $\Omega_b^-$ & $-4/3$ & $0$ & $6084.87\pm9.85$ & $6045.8\pm0.8$ &
                                                                    $39.07\pm9.88$
\\
\hline
\multirow{6}{*}{$\bm 6_{3/2}$}
& $\Sigma_b^{*+}$ & \multirow{3}{*}{$2/3$} & \multirow{3}{*}{$1$} &
     $5825.46\pm9.02$ & $5830.32\pm0.27$ & $-4.86\pm9.03$
\\
& $\Sigma_b^{*0}$ & & & $5827.60\pm9.02$ & --- & ---
\\
& $\Sigma_b^{*-}$ & & & $5831.66\pm9.02$ & $5834.74\pm0.30$ &
                                                              $-3.08\pm9.03$
\\
& $\Xi_b^{*0}$ & \multirow{2}{*}{$-1/3$} & \multirow{2}{*}{$1/2$} &
      $5961.66\pm8.82$ & $5952.3\pm0.6$ & $9.36\pm8.84$
\\
& $\Xi_b^{*-}$ & & & $5965.72\pm8.82$ & $5955.5\pm0.4$ &
                                                         $10.22\pm8.82$
\\
& $\Omega_b^{*-}$ & $-4/3$ & $0$ & $6099.77\pm9.80$ & --- & ---
\\
\hline
\end{tabular}
\end{table*}

\bibliography{DHB}

\end{document}